\documentstyle[prd,manuscript,aps,epsf,floats,amstex]{revtex}
\newcommand{\w}{{@,@,@,@,@,}}

\newcommand{\N}{{\Bbb N}}
\newcommand{\R}{{\Bbb R}}
\newcommand{\s}{{\Bbb S}}
\newcommand{\Z}{{\Bbb Z}}

\newcommand{\Kc}{{\cal K}}

\newcommand{\Mc}{{\cal M}}
\newcommand{\Qc}{{\cal Q}}
\newcommand{\Rc}{{\cal R}}

\newcommand{\Uc}{{\cal U}}
\newcommand{\Vc}{{\cal V}}
\newcommand{\Xc}{{\cal X}}
\newcommand{\Yc}{{\cal Y}}

\newcommand{\Qct}{{\widetilde{{\cal Q}}}}

\newcommand{\Uct}{{\widetilde{{\cal U}}}}
\newcommand{\Vct}{{\widetilde{{\cal V}}}}
\newcommand{\Xct}{{\widetilde{{\cal X}}}}
\newcommand{\Yct}{{\widetilde{{\cal Y}}}}

\newcommand{\Rcb}{{\overline{{\cal R}}}}

\newcommand{\etat}{{\tilde{\eta}}}
\newcommand{\gt}{{\tilde{g}}}
\newcommand{\hti}{{\tilde{h}}}
\newcommand{\Kt}{{\widetilde{K}}}

\newcommand{\Omt}{{\widetilde{\Omega}}}

\newcommand{\qt}{{\tilde{q}}}

\newcommand{\ut}{{\tilde{u}}}
\newcommand{\vt}{{\tilde{v}}}
\newcommand{\xt}{{\tilde{x}}}
\newcommand{\yt}{{\tilde{y}}}

\newcommand{\be}{\begin{eqnarray}}
\newcommand{\ee}{\end{eqnarray}}

\begin{document}
\preprint{Imperial/TP/96-97/55, hep-th/9707177}
  \title{Thermal radiation in non-static curved spacetimes: quantum 
         mechanical path integrals and configuration space topology}
  \author{M.~E.~Ortiz and F.~Vendrell\thanks{Supported by the
   Soci\'et\'e Acad\'emique Vaudoise and by the Swiss National
   Science Foundation.}} 
  \address{Blackett Laboratory, Imperial College, \\ 
  Prince Consort Road, London SW7 2BZ, U.K.}
  \date{June 15, 1998} 
  \maketitle

\begin{abstract}
  A quantum mechanical path integral derivation is given of a thermal 
  propagator in non-static Gui spacetime. The thermal nature of the 
  propagator is understood in terms of homotopically non-trivial paths 
  in the configuration space appropriate to tortoise coordinates. The
  connection to thermal emission from collapsing black holes is discussed. 
\end{abstract}


\newpage

\section{Introduction}

Path integrals can be very intuitive tools with which to tackle certain
problems in quantum mechanics \cite{FHS}. In the context of quantum
field theory, a time-ordered two-point correlation function can be
expressed as a sum over particle paths in a suitable configuration
space. The usefulness of this path integral formalism stems in part
from the fact that it incorporates the physically relevant boundary
conditions encoding state information, and so one need not face the
crucial problem of extracting the adequate solution from the
propagator equation \cite{WMN}. The flip side to this is that a path
integral is not complete without a careful specification of the set of
paths that must be summed over.

A clear advantage of the path integral formalism is that the
dependence on the homotopic properties of the configuration space are
explicit.  Many physical phenomena, such as the Aharonov-Bohm effect,
can be understood to be
a consequence of these properties.  Often in these cases, it is
useful to work with configuration spaces that may have different
topologies to that of the true physical space \cite{SD}.

For an equilibrium situation, the thermal properties of a propagator
are associated via the KMS condition \cite{kms} to periodicity in
imaginary time. This periodicity may be enforced by expressing a
thermal propagator as an infinite sum of non-thermal propagators whose
time arguments are shifted as $t \to t+i\nu\hbar/(kT)$ where $\nu\in\Z$
($\hbar$ and $k$ are the Plank and Boltzmann constants).
Formally (up to the factor of $i$), this kind of expression for a
thermal propagator bears a close resemblance to the propagator of a
particle moving in a multiply connected configuration space. This
relationship has been previously discussed in the context of Euclidean
space path integrals 
(see for example \cite{troost}). The purpose of our work is to provide
a further link between these two ideas by showing that the homotopic
properties of the configuration space of a particle in curved
spacetimes may be regarded as different in a {\it Lorentzian} context, in
coordinates related by a non-analytic transformation.  We argue that
this notion of configuration space topology is closely related to the
thermal properties of non-equilibrium spacetimes. We illustrate these
ideas by focussing on the thermal properties of a non-static spacetime
that can be regarded as a dynamical version of Gui's $\eta-\xi$
spacetime \cite{gui}.

The classic example of the path integral computation of a thermal
propagator is that of the Hartle-Hawking propagator for a black hole
in Schwarzschild coordinates \cite{HH}. Upon Euclideanising the
Schwarzschild metric, the condition that the spacetime be free of
singularities requires that the time coordinate be periodic. The
periodicity leads to a thermal propagator in the obvious way. A
similar conclusion can be reached in a variety of other examples after
an appropriate definition of Euclidean coordinates, such as Rindler
space \cite{troost} and the $\eta-\xi$ spacetime of Gui \cite{gui}. In
some of these examples, it has been shown that the propagator in
Euclidean space can be decomposed according to the winding of paths
around a point in the Euclidean section, and that this decomposition
relates the thermal propagator to a sum of non-thermal propagators
\cite{troost,gui}. This decomposition can be understood in terms of
homotopy if we redefine the Euclidean configuration space by removing
a point.

In this paper we shall apply the path integral formalism to the
thermal emission of particles in a toy model that is analogous to a
collapsing black-hole spacetime.  A collapsing black hole is generally
treated by using the equivalence of a post-collapse hole to an eternal
hole, with the additional information that the matter fields are in
the Unruh vacuum at the past horizon of the eternal hole \cite{Ha}. By looking
at the much simpler example of a non-static Gui spacetime, we directly
develop a notion of configuration space topology for different
coordinates in the Lorentzian signature metric.  We focus on the
different homotopic properties of the configuration space of a
particle viewed in either Kruskal or tortoise type coordinates in the
collapse geometry.  We argue that the thermal nature of the non-static
Gui spacetime can be seen to follow from the non-trivial topology of
the configuration space defined by tortoise coordinates if they are to
cover the entire spacetime. We argue that the appropriate propagator
for thermal emission is indeed obtained by considering paths that
probe this non-trivial topology.  This principle is demonstrated by
exact calculations.

\section{Covariant path integrals}

\subsection{The configuration space}

Let us consider the general case of a particle moving freely on a
curved spacetime $(\Mc,g)$.  We assume that this may be parametrised
entirely with a system of coordinates denoted by $q$, whose values
belong to a connected manifold $\Qc$, the {\it configuration space},
endowed with the metric $g_q$.  A virtual path followed by a particle
in $\Mc$ is given in theses coordinates by the function 
$q=q(\omega)$, where $\omega$ is the parameter-time. This path is denoted 
by $[\w q\w]$.  This function is continuous everywhere except maybe
at some values of $\omega$, because the coordinates $q$ may not cover
$\Mc$ in a continuous way.  In a relativistic framework, the particle is
allowed to go forward and/or backwards in the time-coordinate $q^0$.

We shall be interested in the case where the configuration space $\Qc$
is multiply connected \cite{LL}. Normally, a multiplicity of paths connecting a
pair of given points in $(\Mc,g)$ is associated with the topology of
the manifold $\Mc$. However, since the paths take values in $\Qc$, it
is possible that similar effects can arise when $(\Mc,g)$ is
parametrised in a coordinate system that is related non-analytically
to a given coordinate patch of $\Mc$. We shall see this explicitly
below.
 
When the manifold $(\Qc,g_q)$ is multiply connected, it is useful to
consider its {\it covering space} $(\Qct,\gt_\qt)$ \cite{LL}.  In this
case, $\Qct$ contains many distinct {\it images} of any given point of
the configuration space.  The elements of the set $\Gamma$ are defined
to be {\it isometries} which relate these images, so that one has
$\Qc=\Qct/\Gamma$.  The images of $q\in\Qc$ will be denoted by
$\qt^\nu\in\Qct$, where $\nu$ ranges over the elements of $\Gamma$.
The configuration space is recovered from its connected covering space
when all these images are identified.  The metric $\gt$ in $\Qct$ is
defined naturally by $\gt(\qt^\nu)=g(q)$.

Similarly, a path $[\w q\w]$ in $\Qc$ has different images in $\Qct$.  
These are continuous functions and may be labelled according to the
images of the endpoints.  The base point in $\Qct$ of the endpoint $q_i\in\Qc$
is defined by the particular image point $\qt_i\equiv \qt_i^{\nu=0}$.
The paths $[\w\qt\w]$ with endpoints $(\qt_i,\qt_f^\nu)$ belong to a given
homotopy class.  Paths of different classes may not be continuously
deformed into one another.  The set of all classes combined with a
composition operator defines the fundamental group of $\Qct$.  

\subsection{The propagator and the path integral}

In quantum mechanics, the amplitude to move from an initial point
$q_i\in\Qc$ to a final point $q_f\in\Qc$ in a parameter-time
$s=s_f-s_i$ is given by the propagator $K_q(q_i,q_f;s)$, or heat kernel,
in $q$ cordinates. It does not depend on the mass of the particle and 
satisfies the Schr\"odinger equation \cite{BP}
\be
i\hbar\w\partial_s\w K_q\w(q',q'';s)=
-\hbar^2\w\nabla^\mu\nabla_\mu\w K_q\w(q',q'';s),
\label{Schr}
\ee
where $\nabla_\mu$ is the covariant derivative.
This propagator is related the usual propagator $G_q\w(q_i,q_f)$ of
quantum field theory by a Fourier transform in $s$ whose conjugate
variable is $m^2$.

Because it satisfies the Schr\"odinger equation (\ref{Schr}), the
propagator $K_q\w(q_i,q_f;s)$ can be written as a sum over paths joining
the endpoints $q_i$ and $q_f$ within a parameter time $s$. 
In $q$ coordinates, the sum is taken over the paths contained within
$\Qc$, and is written symbolically as
\be
K^{vac}_q\w(q_i,q_f;s) =
\sideset{}{_{g_q}}\sum_{_{[\w q\w]\,\in\,\Qc}^{q_i\rightarrow q_f}}
\exp\left(\w \frac{i}{\hbar}\w S_{g_q}\w[\w q\w]\w\right),
\label{SPq}
\ee
where the covariant action $S_{g_q}\w[\w q\w]$ is given by
\be
S_g\w[\w q\w] = 
\frac{1}{4}\int_{s_i}^{s_f}\,d\omega\,g_{\mu\nu}(q)\,\dot{q}^\mu\,\dot{q}^\nu.
\ee
This sum over paths is defined below. With this definition,
the propagator $K^{vac}_q\w(q_i,q_f;s)$ is a {\it particular} solution
of the Schr\"odinger equation (\ref{Schr}). The choice of its boundary 
conditions is equivalent to the choice of the configuration space
$\Qc$. The latter choice is natural since it is fixed by the manifold
$\Mc$, or spacetime itself, and by the coordinates $q$. The set of paths
on which the sum is taken defines actually a vacuum since, in quantum
field theory, the Fourier transform $G^{vac}_q\w(q_i,q_f)$ of the 
propagator (\ref{SPq}) is the scalar two-point correlation function in 
this vacuum.  

The sum over paths (\ref{SPq}) is calculated in the covering space 
$\Qct$. It is rewritten as sum over paths in $\Qct$ by taking into account
its multiply connected topology, i.e.~by summing over the classes of paths, 
or equivalently over the images $\qt_f^\nu$,
\be
\sideset{}{_{g_q}}\sum_{^{q_i\rightarrow q_f}_{[\w q\w]\,\in\,\Qc}} 
\exp\left(\w\frac{i}{\hbar}\w S_{g_q}\w[\w q \w] \w \right) =
\sum_{\nu}\ \sideset{}{_{\gt_\qt}}
\sum_{_{[\w \qt\w]\,\in\,\Qct}^{\qt_i\rightarrow\qt_f^\nu}} 
\exp\left(\w\frac{i}{\hbar}\w S_{\gt_\qt}\w[\w \qt \w] \w \right).
\label{Pathint1}
\ee
On the r.h.s.~of this equation the sum is taken over paths joining the 
endpoints $\qt_i$ and $\qt_f^\nu$. It is now natural to define a propagator 
$\Kt_\qt$ in $\Qct$ by the sum over paths
\be
\Kt_\qt\w(\qt';\qt'';s)= \sideset{}{_{\gt_\qt}}
\sum_{^{\,\qt'\rightarrow \qt''}_{[\w \qt\w]\,\in\,\Qct}}
\exp\left(\w\frac{i}{\hbar}\w S_{\gt_\qt}\w[\w \qt \w] \w \right).
\ee
Since the set of paths over which the sum is taken in this definition is 
formally different from that of Eq.~(\ref{SPq}), this propagator defines a 
new vacuum $vac'$. Going back to the configuration space $\Qc$, if one 
defines the propagator $K^{vac'}_q$ through the identification 
$\Kt_\qt\equiv K^{vac'}_q$, one obtains the result \cite{Ma}
\be
K^{vac}_q\w(q_i;q_f;s)=\sum_{\nu}K^{vac'}_q\w(\qt_i;\qt_f^\nu;s),
\label{sumnu}
\ee
which is true in all multiply connected manifolds.

Practically, a sum over paths is calculated from a path integral. One can
show that this is given by \cite{BP}
\be
\sideset{}{_{\gt_\qt}}
\sum_{^{\,\qt'\rightarrow \qt''}_{[\w \qt\w]\,\in\,\Qct}}
\exp\left(\w\frac{i}{\hbar}\w S_{\gt_\qt}\w[\w \qt \w] \w \right) =
\underset{[\w \qt\w]\,\in\,\Qct}{\int_{\qt'}^{\qt''}} D_{\gt_\qt}\w[\w\qt\w]\, 
\exp\left(\w\frac{i}{\hbar}\w S_{\gt_\qt}\w[\w\qt\w]\,\right).
\label{Pathint}
\ee
If the scalar curvature vanishes, the path integral on the r.h.s.~of
this equation is defined by
\begin{eqnarray}
\underset{[\w\qt\w]\,\in\,\Qct}{\int_{\qt'}^{\qt''}} \!
D_{\gt_\qt}\w[\w\qt\w] \ e^{\frac{i}{\hbar}\w S_{\gt_\qt}\w[\w\qt\w]} = 
\lim_{N\rightarrow\infty} \left(\frac{N}{4\pi\hbar is}\right)^{2N}
\prod_{j=1}^{N-1}
\int_{\Qct} d^4\qt_j \parallel\!\gt_\qt(\qt_j)\!\parallel^{\frac{1}{2}} 
e^{\frac{i}{4\hbar}\w\sum_{k=1}^N\int_{s_{k-1}}^{s_k} \!\!
d\omega \,\gt_{\mu\nu}\w\dot{\qt}^\mu(\omega)\w\dot{\qt}^\nu(\omega)}
\label{defPI} 
\end{eqnarray}
where $s_j= s_i+js/N$, $\qt_j=\qt(s_j)$ ($j=0,1,...,N$), $N\in\N$ and 
$\parallel\ \parallel$ denotes the absolute value of the determinant. Each
integral in the exponential is evaluated along the image-{\it geodesic} 
connecting $\qt_{j-1}$ and $\qt_j$ which ensures that the path integral
is covariant.  

\section{A simple model}

\subsection{The non-static Gui spacetime}

\begin{figure}
\centerline{\epsfysize=3.0truein\epsfbox{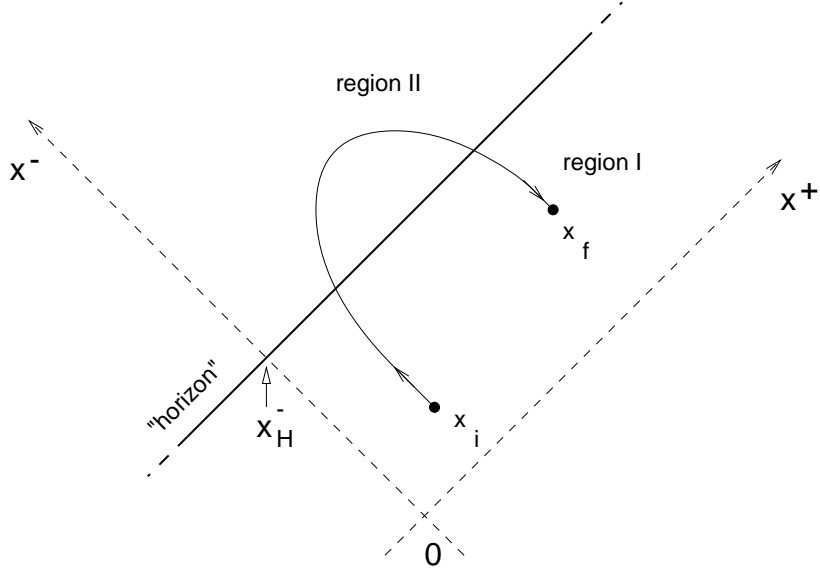}}
\vspace{5mm}
\caption{Spacetime diagram of the Gui non-static spacetime. A path crossing
twice the ``horizon'' is shown.} 
\label{fig:gui}
\end{figure}

Let us now introduce a simple dynamical spacetime model that will
allow us to perform explicit calculations.  This two-dimensional model
is defined by the line element \cite{FV}
\begin{eqnarray}
ds^2_h=\frac{dx^+\,dx^-}{\kappa\left(x^-_H-x^-\,\right)},
\label{Gds}
\end{eqnarray}
where $x^\pm=x^0\pm x^1$, $x^-_H\in\R$ and $\kappa>0$. The $x$
coordinates are Kruskal-like coordinates. This line element
is the dynamical equivalent of the eta-xi spacetime of Gui\footnote{The
static Gui's metric is defined in four dimensions by 
$ds^2_{Gui}=\kappa^{-2}\frac{dx^+\,dx^-}{x^+ x^-} - (dx^2)^2 -(dx^3)^2$ 
where $x^\pm=x^0\pm x^1$.} \cite{gui}.
We define regions $I$ and $I\!I$ by the half-planes $x^-<x^-_H$ and 
$x^->x^-_H$ respectively, see Fig.~\ref{fig:gui}.
Two sets of tortoise-like coordinates $y_I$ and $y_{I\!I}$ are defined there
by\footnote{ The multiplicative factor in front of the exponential and
the additive factor inside of it may in fact be arbitrary. The
motivation of the present choice will become clear in subsection 
\ref{subsec:Sch} when the collapsing Schwarzschild black hole will be
considered. }
\begin{eqnarray}
x^-(y^-_I) & = & x^-_H - 2\exp\left[-\kappa\w(y^-_I-x^-_H)\w\right],
\label{x(y1)}  \\ [2mm]
x^-(y^-_{I\!I}) & = & x^-_H + 2\exp\left[-\kappa\w(y^-_{I\!I}-x^-_H)\w\right],
\label{x(y2)}
\end{eqnarray}
and by $x^+(y^+_{I,I\!I}) = y^+_{I,I\!I}$.
These new coordinates are actually Minkowski coordinates since we have
\begin{eqnarray}
ds^2_h = dy^+_a\,dy^-_a, \qquad a=I,I\!I.
\end{eqnarray}
This means that the spacetime we have defined is made of two causally
and classically disconnected copies of Minkowski spacetime glued
together.
The null line $x^-=x^-_H$ will be called the ``horizon''.

For this simple model, the configuration spaces $\Xc$ and $\Yc_a$ 
($a=I,I\!I$) in $x$ and $y$ coordinates are given by
\begin{eqnarray}
\Xc &=& \{\,(x^+,x^-)\in\R^2\,\}, \\ [2mm]
\Yc_a &=& \{\,(y^+_a,y^-_a)\in\R^2\,\},
\qquad a=I,I\!I,
\label{Yca}
\end{eqnarray}
and these are isomorphic to their corresponding covering spaces
$\Xct$ and $\Yct_a$, $a=I,I\!I$.

\subsection{The complex covering space and its topology}

The utility of the coordinates $y_a$ ($a=I,I\!I$) introduced
in the preceding section stems from the fact that their are Minkowskian.
It then seems possible to calculate explicitly the path integral in these
coordinates.
However, to actually calculate a sum over paths it is necessary to work
with a connected configuration space, instead for example of the two
sets $\Yc_I$ and $\Yc_{I\!I}$, see Eq.~(\ref{Yca}). One needs to be able
to parametrise a path crossing the ``horizon" with only one set of 
coordinates. 

\begin{figure} 
\centerline{\epsfysize=2.5truein\epsfbox{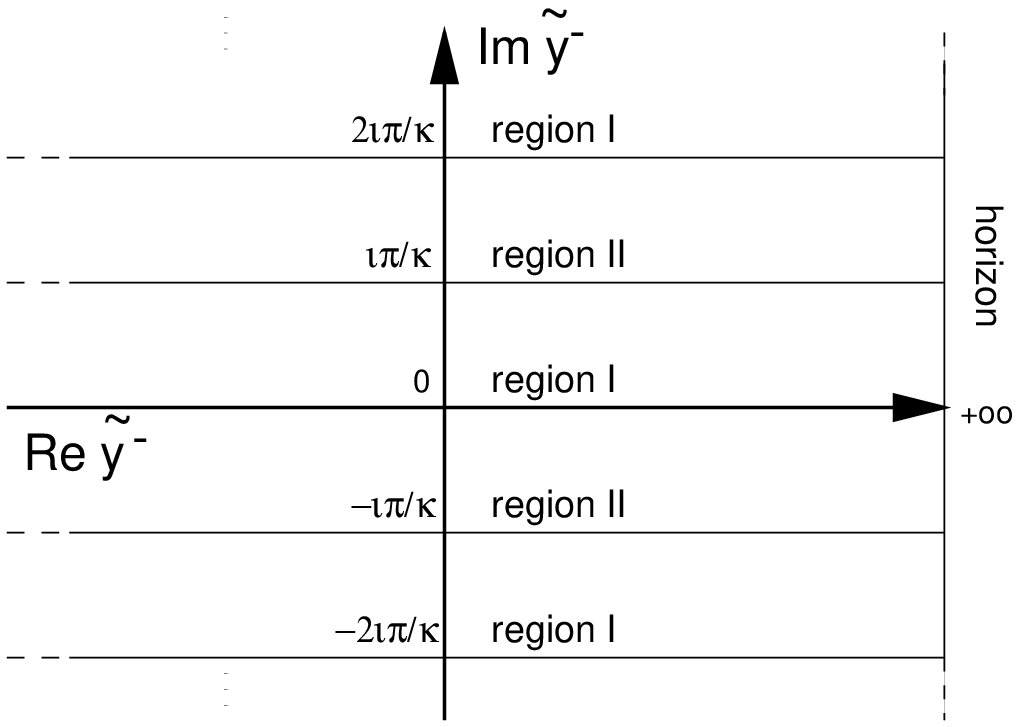}}
\vspace{5mm}
\caption{A section of the tortoise complex covering space $\Yct$ along the 
$\yt^-$ coordinate.}
\label{fig:guiconf}
\end{figure}

Since one does not have a {\it real} set of Minkowski coordinates to
cover both the regions $I$ and $I\!I$ at our disposal, we shall
introduce a {\it complex} set of Minkowski coordinates denoted by $y$.
These are defined in the covering space by the identifications
\begin{eqnarray}
\yt^-_I &\equiv& \yt^- +i2\pi\nu/\kappa, 
\label{yequiv1}\\ [2mm]
\yt^-_{I\!I} &\equiv& \yt^- +i2\pi(\nu+1/2)/\kappa,
\label{yequiv2}
\end{eqnarray}
where $\nu\in\Z$, and $\yt^+_{I,I\!I}=\yt^+$.
The transformations (\ref{x(y1)}) and (\ref{x(y2)}) may then be written 
together in a compact form:
\begin{eqnarray}
\xt^-(\yt^-) =  x^-_H - 2\exp\left[-\kappa\w(\yt^--x^-_H)\w\right].
\label{x(y)c}
\end{eqnarray}
The necessity of allowing arbitrary integer values for $\nu$ in 
Eqs.~(\ref{yequiv1}) and (\ref{yequiv2}) stems from the fact that one needs to 
parametrise continuously a path crossing for example several times the
``horizon" as shown below. The complex covering space $\Yct$ is then given by
\begin{eqnarray}
\Yct = \{\,(\yt^+,\yt^-)\in 
\R\times\bigcup_{\mu\,\in\,\Z} A_{\mu/2}\bigcup B \,\},
\label{AB}
\end{eqnarray}
where $A_\mu=\R+i2\pi\mu/\kappa$ and $B = +\infty + i \R$. The set $B$
parametrises the ``horizon" and the sets $\bigcup_{\mu\,\in\,\Z} A_{\mu}$ and
$\bigcup_{\mu\,\in\,\Z} A_{\mu+1/2}$ the regions $I$ and $I\!I$ 
respectively. The space $\Yct$ is represented in Fig.~\ref{fig:guiconf}.

\begin{figure}
\centerline{\epsfysize=1.6truein\epsfbox{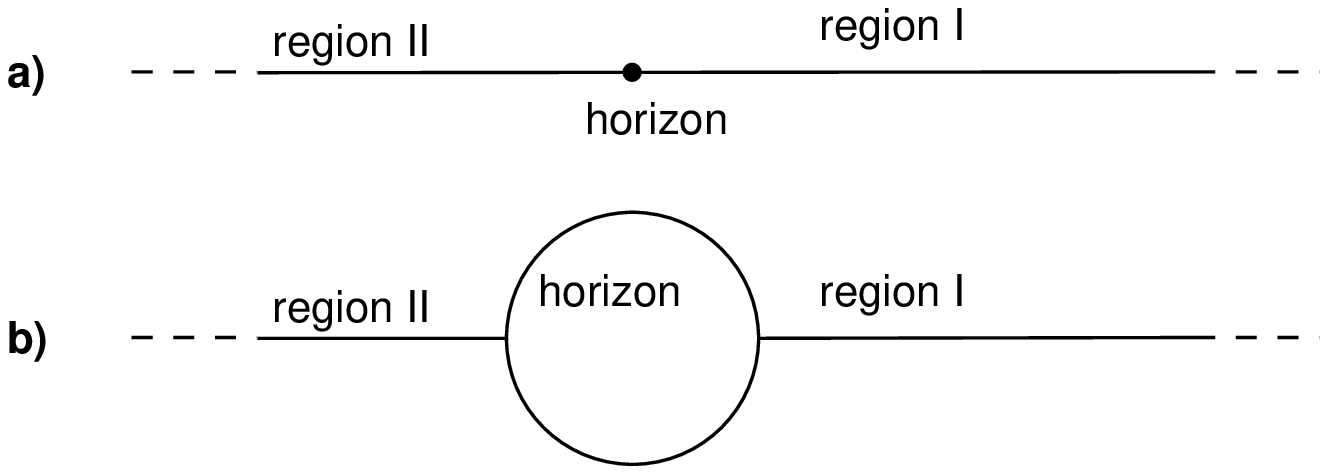}}
\vspace{5mm}
\caption{A section of the configuration spaces a) $\Xc$ and b) $\Yc$
along the $x^-$ and $y^-$ coordinates respectively. }
\label{fig:guitopo}
\end{figure}

A point $y$ in $\Yc$ has thus a denumerable infinite number of images
in $\Yct$. These will be denoted by $\yt^\nu$, where $\nu\in\Z$, and are
defined in Eqs.~(\ref{yequiv1}) and (\ref{yequiv2}) if the point $y$ is located
in the regions $I$ and $I\!I$ respectively.
These images are related through an isometry of the set
\begin{eqnarray}
\Gamma &=&\{\,\gamma_\nu: (\yt^+,\yt^-) \mapsto 
(\yt^+,\yt^-+i2\pi\nu/\kappa),\, \nu \in\Z\,\}.
\end{eqnarray}
The configuration space $\Yc$ is obtained by identifying them in such
a way that $\Yc=\Yct/\Gamma$. The resulting manifold is shown in
Fig.~\ref{fig:guitopo}.
Clearly, it is not simply connected. The topology of the ``horizon" in
the configuration space in the coordinates $(y^+,y^-)$ is thus that of
$\R\times\s^1_{1/\kappa}$, where $\s^1_{1/\kappa}$ is a
circle of radius $\kappa^{-1}$. The fact that the topologies of $\Xc$ and
$\Yc$ are different stems from the fact that the transformation (\ref{x(y)c})
is not analytic.

\begin{figure}
\centerline{\epsfysize=2.0truein\epsfbox{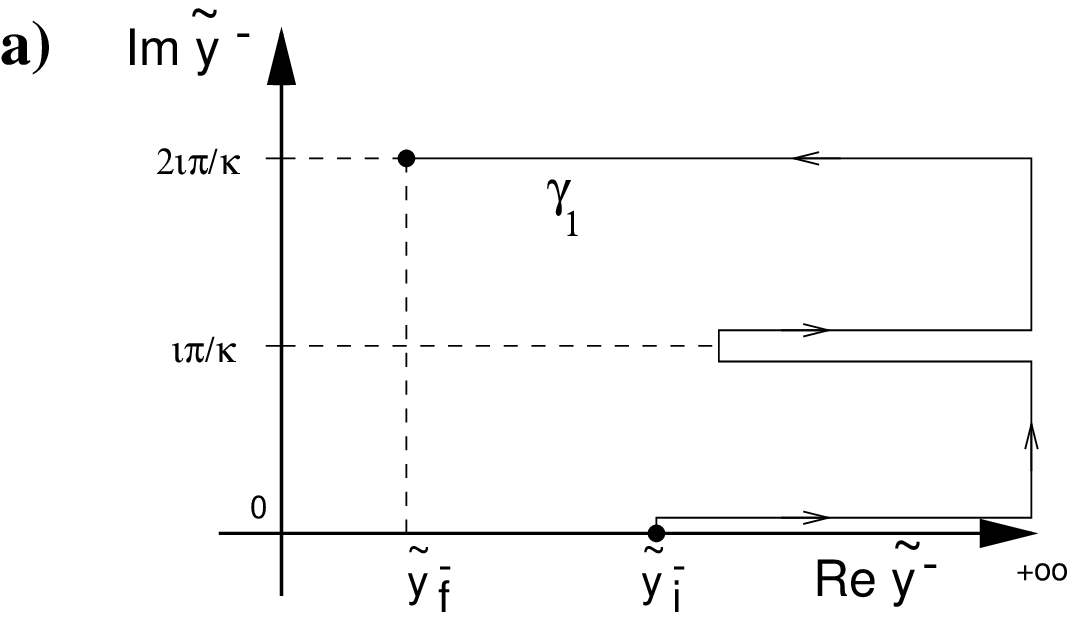}}
\centerline{\epsfysize=1.2truein\epsfbox{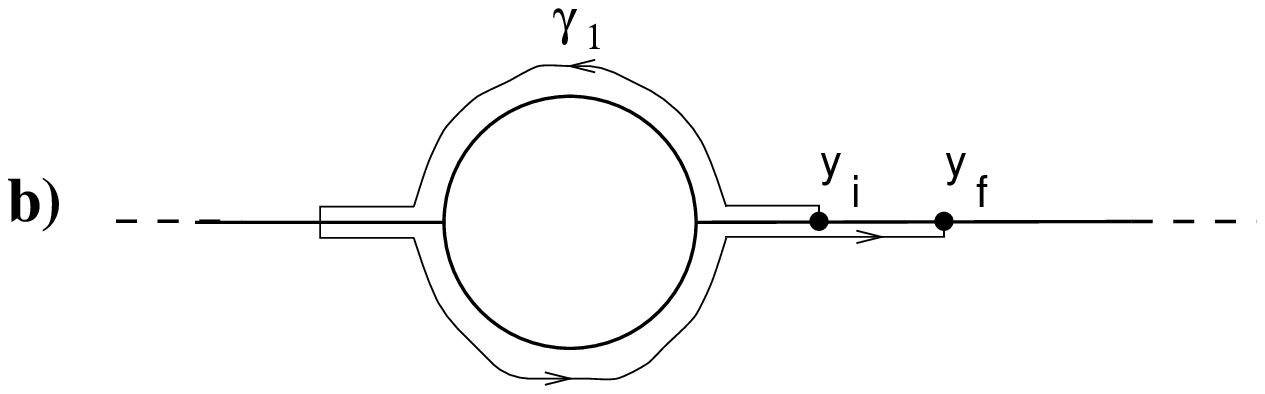}}
\vspace{5mm}
\caption{A path crossing the horizon twice with winding number +1 in the
a) covering space $\Yct$ and b) in the configuration space $\Yc$.}
\label{fig:path1}
\end{figure}

\begin{figure}
\centerline{\epsfysize=2.0truein\epsfbox{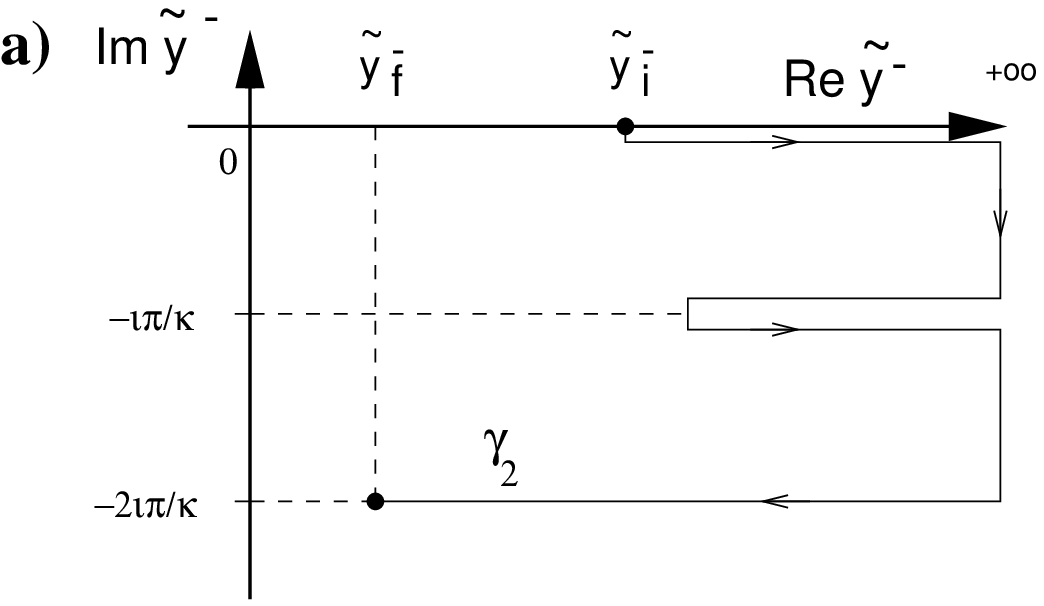}}
\centerline{\epsfysize=1.2truein\epsfbox{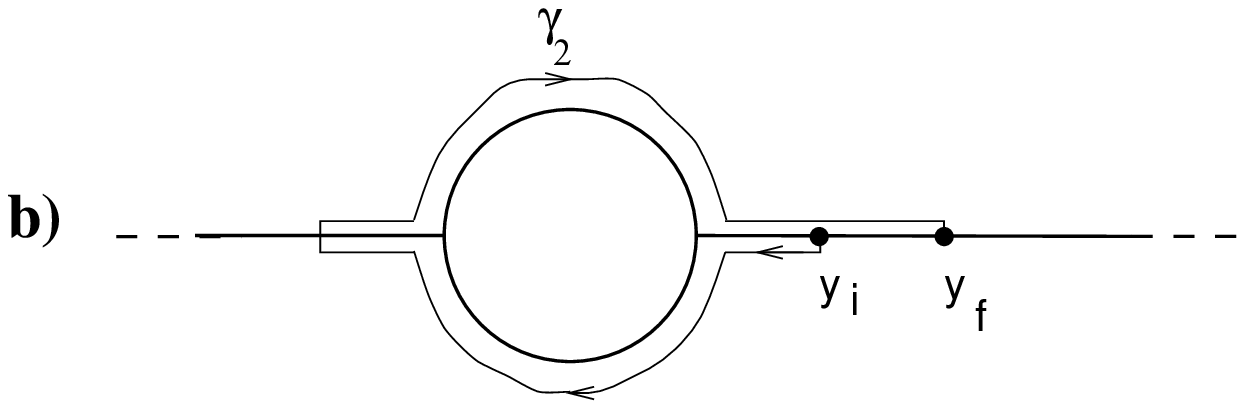}}
\vspace{5mm}
\caption{A path crossing the horizon twice with winding number -1 in the
a) convering space $\Yct$ and b) in the configuration space $\Yc$. }
\label{fig:path2}
\end{figure}

\begin{figure}
\centerline{\epsfysize=1.6truein\epsfbox{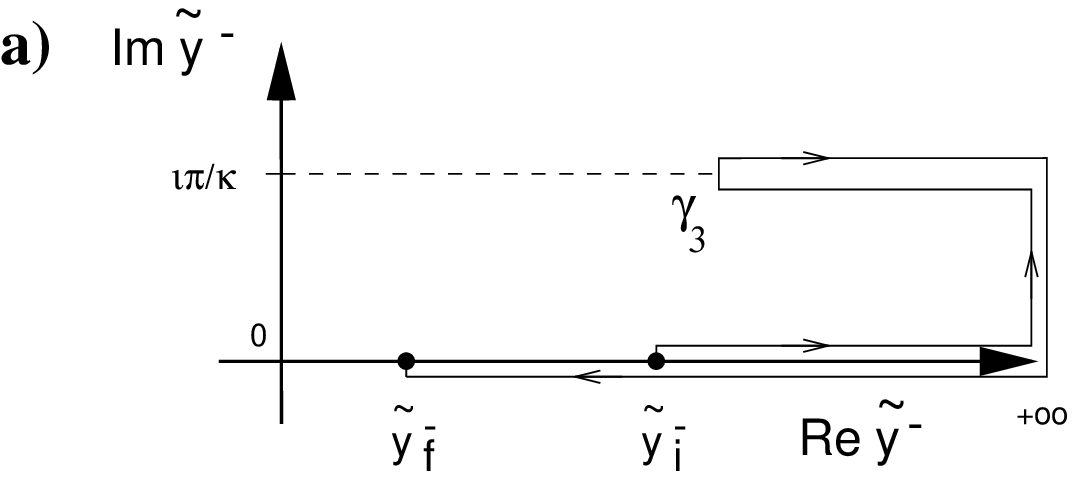}}
\centerline{\epsfysize=1.2truein\epsfbox{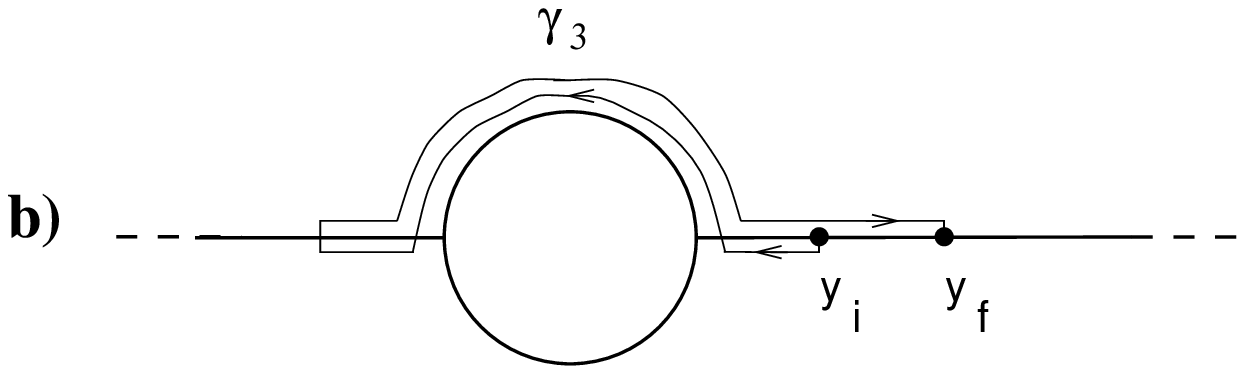}}
\vspace{5mm}
\caption{A path crossing the horizon twice with winding number 0 in the
a) covering space $\Yct$ and b) in the configuration space $\Yc$.}
\label{fig:path3}
\end{figure}

The fundamental group of $\Yc$ is thus isomorphic to $\Z$.  The classes 
of paths are labelled by the integer $\nu$ which is the winding number of the 
paths around the circle $\s^1_{1/\kappa}$.  A path crossing the ``horizon" 
twice is represented in Fig.~\ref{fig:gui}.  Their images of winding
number $+1$, $-1$ and $0$ in the covering and configuration spaces are shown in
Figs.~\ref{fig:path1}, \ref{fig:path2} and \ref{fig:path3} respectively.

\subsection{Applying the path integral formalism}

We now turn to the problem of calculating the propagator from the path
integral formalism. This is expressed as a sum over paths in the configuration 
space $\Xc$,
\be
K^{in}_x(x_i,x_f;s) = \sideset{}{_{h_x}}
\sum_{^{x_i\rightarrow x_f}_{[\w x\w]\,\in\,\Xc}}
\exp\left(\w\frac{i}{\hbar}\w S_{h_x}\w[\w x\w]\w\right),
\label{SPx}
\ee
where the metric $h_x$ in $x$ coordinates is given in Eq.~(\ref{Gds}).
This expression defines the incoming vacuum $in$.
It is difficult to calculate the sum over paths in this form, because
of the presence of the non-trivial metric $h_x$.
Our strategy is to transform it to the complex $y$ coordinates 
covering the entire spacetime, in which the metric is Minkowskian.
The relevant connected configuration space to consider is then $\Yc$. 
The interesting physics follows from the non-trivial topology
of this configuration space as we show now.

The sum over paths (\ref{SPx}) when written as a path integral in the 
covering space $\Xct$ is badly defined because the metric $h_x$ 
is singular at $x^-=x^-_H$. The obvious way to define it correctly is
to performed the integration over $\xt^-$ using the principal value.
This is thus defined by
\begin{eqnarray}
\sideset{}{_{h_x}}\sum_{^{x_i\rightarrow x_f}_{[\w x\w]\,\in\,\Xc}}
\exp\left(\w\frac{i}{\hbar}\w S_{h_x}\w[\w x\w]\w\right) 
= \underset{[\w x\w]\,\in\,\Xct^*}{\int_{\xt_i}^{\xt_f}} 
D_{\hti_\xt}\w[\w\xt\w]\, 
\exp\left(\w\frac{i}{\hbar}\w S_{\hti_\xt}\w[\w\xt\w]\w\right),
\label{PIx*}
\end{eqnarray}
where the space $\Xct^*$ is given by
\begin{eqnarray}
\Xct^* &=& \{\,(\xt^+,\xt^-)\in\R\times\R^*\,\},
\end{eqnarray} 
if $\R^*=\lim_{\epsilon\rightarrow{0}}\,(-\infty,-\epsilon]
\cup[+\epsilon,+\infty)$. 
Interestingly, this definition amounts to getting rid of the paths crossing 
the ``horizon" at one or more of the parameter-time values $s_j$ of
Eq.~(\ref{defPI}), where $j=1,...,N-1$ ($N\in\N$).

We now perform a change of coordinates in the sum over paths
and because the propagator is a biscalar one may write
\be
\sideset{}{_{h_x}}\sum_{^{x_i\rightarrow x_f}_{[\w x\w]\,\in\,\Xc}}
\exp\left(\w\frac{i}{\hbar}\w S_{h_x}\w[\w x\w]\w\right) =
\sideset{}{_\eta}\sum_{^{y_i\rightarrow y_f}_{[\w y\w]\,\in\,\Yc} }
\exp\left(\w\frac{i}{\hbar}\w S_\eta\w[\w y\w]\w\right).
\ee
Since the sum over paths on the r.h.s.~is also badly defined when written as 
a path integral in the covering space $\Yct$, we introduce
the space $\Yct^*$ by
\be
\Yct^* = \{\,(\yt^+,\yt^-)\in\R\times\bigcup_{\mu\,\in\,\Z} A_{\mu/2}\,\},
\ee
i.e.~we removed the set $\R\times B$ from $\Yct$, and we
rewrite the sum over paths in the form
\be
\sideset{}{_\eta}\sum_{^{y_i\rightarrow y_f}_{[\w y\w]\,\in\,\Yc}}
\exp\left(\w\frac{i}{\hbar}\w S_\eta\w[\w y\w]\w\right) = 
\sum_{\nu\in\Z} \underset{[\w\yt\w]\,\in\,\Yct^*}{\int_{\yt_i}^{\yt^\nu_f}} 
D_\etat\w[\w\yt\w]\,\exp\left(\w\frac{i}{\hbar}\w S_\etat\w[\w\yt\w]\w\right),
\label{pii}
\end{eqnarray}
where $\etat=\eta$. In this last equation, we have taken into account the 
multiply connected nature of $\Yc$ by summing over the different classes of 
path with winding number $\nu$.
Each integral in the sum (\ref{pii}) of path integrals is given by
\be
\int_{\Yct^*} \!\! d^2\yt_j \
e^{\frac{i}{4\hbar} \sum_{k=1}^N\int_{s_{k-1}}^{s_k}\!\!
d\omega\,\etat_{\mu\nu}\,\dot{\yt}^\mu(\omega)\,\dot{\yt}^\nu(\omega)}
&=& \sum_{\mu\in\Z}\int_{A_{\mu/2}}\!\!\!\!d\yt^-_j\int_{\R}\!\!\!d\yt^+_j\,
e^{\frac{i}{4\hbar} \sum_{k=1}^N\int_{s_{k-1}}^{s_k} \!\!
d\omega \, \etat_{\mu\nu}\,\dot{\yt}^\mu(\omega)\,\dot{\yt}^\nu(\omega)}.
\ee
Since the integrand in the r.h.s.~of this last equation does not actually 
depend on the imaginary value of $\yt^-_j$, we have
\be
\int_{\Yct^*} d^2\yt_j \
e^{\frac{i}{4\hbar} \sum_{k=1}^N\int_{s_{k-1}}^{s_k}\!
d\omega \, \etat_{\mu\nu}\,
\dot{\yt}^\mu(\omega)\,\dot{\yt}^\nu(\omega)}
= C \int_{\R^2} d^2\yt_j\
e^{\frac{i}{4\hbar} \sum_{k=1}^N\int_{s_{k-1}}^{s_k}\!d\omega\, 
\etat_{\mu\nu}\,\dot{\yt}^\mu(\omega)\,\dot{\yt}^\nu(\omega)},
\end{eqnarray}
where $C$ is an infinite constant, which can be removed by normalising
correctly the path integral to take into account 
the fact that the integration is made over an infinite number of copies
of $\R^2$. The sum over paths becomes then
\begin{eqnarray}
\sideset{}{_{g_x}}\sum_{^{x_i\rightarrow x_f}_{[\w x\w] \,\in\,\Xc}}
\exp \left(\w\frac{i}{\hbar}\w S_{g_x}\w[\w x\w]\w\right) 
&=& \sum_{\nu\in\Z}\underset{[\w\yt\w]\,\in\,\R^2}{\int_{\yt_i}^{\yt^\nu_f}}
D_{\etat}\w[\w\yt\w]\,\exp\left(\w\frac{i}{\hbar}\w 
S_{\etat}\w[\w\yt\w]\w\right).
\end{eqnarray}
It has thus been rewritten as a sum of {\it real}
path integrals, although the final endpoint is complex.  
It is regularised by performing two Wick rotations in both the 
parameter-time and the time-coordinate in the standard way \cite{HO}.  
The contribution of the class of paths with winding number $\nu$ is the 
free propagator $\Kt_0$ in $\yt$ with arguments $\yt_i$ and $\yt_f^\nu$:
\begin{eqnarray}
\underset{[\w\yt\w]\,\in\,\Yct^*}{\int_{\yt_i}^{\yt^\nu_f}}
D_\etat\w[\w\yt\w]\,\exp\left(\w\frac{i}{\hbar}\w S_\etat\w[\w\yt\w]\w\right) 
= \Kt_0\left(\,\yt_i,\,\yt_f^\nu;\,s\,\right).
\label{clnu}
\end{eqnarray}
The outgoing vacuum is defined by $\Kt_0=K^{out}_0$.
By adding all these propagators, the total propagator is obtained
\be
K^{in}_y\left(\,y_i;\,y_f;\,s\,\right)
= \sum_{\nu\in\Z} K^{out}_0
\left(\,y_i;\,y^+_f, y^-_f+i2\pi\nu/\kappa;\,s\,\right).
\label{Gfinal}
\ee
The propagator therefore represents a {\it right-moving} thermal flux of
particles with temperature $T=\hbar\kappa/(2\pi k)$, as can be shown by
calculating the component $(--)$ of the energy-momentum tensor from this
expression \cite{FV}. 

\subsection{Relation to the collapsing Schwarzschild black hole}
\label{subsec:Sch}

\begin{figure}
\centerline{\epsfysize=3.0truein\epsfbox{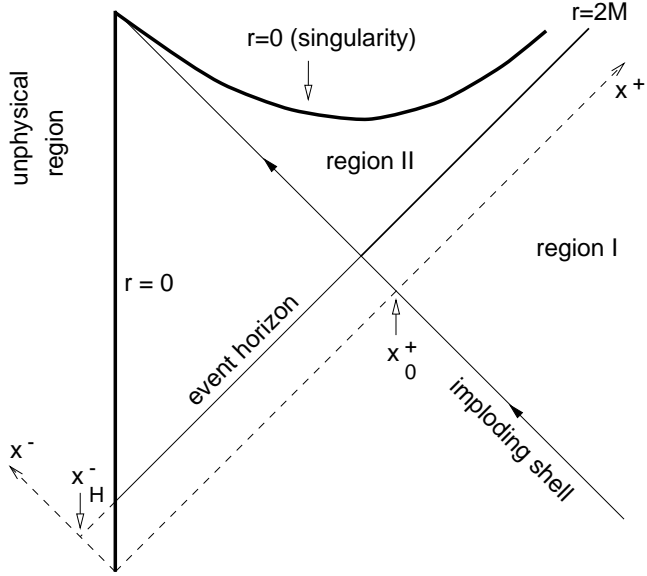}}
\vspace{5mm}
\caption{Spacetime diagram of the Synge collapse in Kruskal coordinates. 
Each point represents a sphere $\s^2$. The physical region, or
accessible region of spacetime, defines the Kruskal configuration space
$\Kc$.}
\label{fig:sch}
\end{figure}

A collapsing Schwarz\-schild black hole can be modelled by an
imploding spherical shell of radiation. If one assumes that the shell is
infinitesimally thin as did Synge \cite{Sy}, the only non-vanishing 
energy-momentum tensor component in the Kruskal coordinates $u$ is
\begin{eqnarray}
T_{++}(u)= \frac{M}{4\pi r^2}\,\delta(u^+-u^+_0),
\label{Tcs}
\end{eqnarray}
where $M>0$ and $u^+_0>0$. From the equations of General Relativity one
obtains the line element \cite{Sy}
\begin{eqnarray}
ds^2_g = \left[\,1-\frac{2m(u^+)}{r}\,\right]\,
\frac{du^+\,du^-}
{\left[\,1-\displaystyle\frac{4m(u^+)}{(u^+_0-u^-)}\,\right]} 
- r(u^+,u^-)^2\,d\Omega^2,
\label{dsx}
\end{eqnarray}
where $d\Omega^2=d\theta^2+\sin^2\theta\,d\phi^2$ and
$m(u^+)=M\,\theta(u^+-u^+_0)$.  Inside the collapsing shell, one has
$u^\pm=t\pm r$, where $t$ is the time coordinate and $r\geq0$ the
radius, and the line element is Minkowskian there.  Outside the
imploding shell, the Kruskal and Schwarzschild coordinates $(t,r)$ are
related implicitly by
\begin{eqnarray}
\exp\left(-2\w\kappa\w t\right)
&= &\vert\,u^-_H-u^-\,\vert\ \exp\left[\kappa\w(u^-_H-u^--u^+)\right], 
\label{ut} \\ [2mm]
2\left(\,r-2M\,\right)\,\exp\left(2\w\kappa\w r\right) &=& 
\left(\,u^-_H-u^-\,\right)\,\exp\left[\kappa\w(u^+-u^-)\right],
\label{ur}
\end{eqnarray}
where $\kappa=1/(4M)$ and $u^-_H=u^+_0-4M$. The collapsing spacetime is
represented in Fig.~\ref{fig:sch}. The horizon is located at $u^-=u^-_H$
or $r=2M$ if $u^+>u^+_0$. The Kruskal configuration space $\Uc$ is given by
\begin{eqnarray}
\Uc=\{\,(u^+,u^-,\theta,\phi)\in\R^2\times\s^2 
\ \vert\ r(u^+,u^-)\geq 0\,\}.
\end{eqnarray}
It is homotopically trivial, i.e.~it is isomorphic to its covering space
$\Uct$.

Two new sets of coordinates $v_I$ and $v_{I\!I}$ can be defined for
$u^-<u^-_H$ and $u^->u^-_H$ respectively by
\begin{eqnarray}
v^-_I(u^-) &=& u^--\kappa^{-1}\,\ln\left[\,(u^-_H-u^-\,)/2\right],
\label{v(u)1} \\ [2mm]
v^-_{I\!I}(u^-) &=& u^--\kappa^{-1}\,\ln\left[\,(u^--u^-_H\,)/2\right],
\label{v(u)2}
\end{eqnarray}
and $v^+_{I,I\!I}(u^+) = u^+$.  The $v_{I,I\!I}$ are tortoise coordinates.
Outside the shell, the line element (\ref{dsx}) is the
Schwarzschild line element in $(t,r)$ coordinates where
$(v^0_{I,I\!I},v^1_{I,I\!I}) =
(t,r+2M\,\ln\left\vert\,r-2M\,\right\vert)$. 

We will be interested in a connected region $\Rc$ defined by 
$u^-\approx u^-_H$ and $u^+\gg1$. Within this region, there is a subregion 
which is located far from the black hole and at late times 
(i.e.~$r\gg 2M$ and $t\gg1$).
This subregion is denoted by $\Rcb$. An inertial observer far from the
black hole is contained within this region at late times.
In $\Rcb$, the line element (\ref{dsx}) becomes
\be
ds^2_g &\approx& 
\frac{du^+\,du^-}{\kappa\left(u^-_H-u^-\,\right)} - r(u^+,u^-)^2\,d\Omega^2,
\label{SdsR}
\ee
and the transformations (\ref{v(u)1}) and (\ref{v(u)2}) are given if
$u^-\approx u^-_H$ by
\be
u^-(v^-_I) &\approx& u^-_H- 2\exp\left[-\kappa\,(v^-_I-u^-_H)\w\right],
\label{u(v)R1} \\[2mm]
u^-(v^-_{I\!I}) &\approx& u^-_H
+ 2\exp\left[-\kappa\,(v^-_{I\!I}-u^-_H)\w\right].
\label{u(v)R2}
\ee
Under the identifications $u\equiv x$ and $v_{I,I\!I}\equiv y_{I,I\!I}$,
these clearly agree formally with the metric (\ref{Gds}) and transformations 
(\ref{x(y1)}) and (\ref{x(y2)}) of the non-static Gui spacetime, if 
abstraction is made of the angular degrees of freedom.

The complex tortoise coordinate $v$ is defined from the coordinates $v_I$
and $v_{I\!I}$ as in Eqs.~(\ref{yequiv1}) and (\ref{yequiv2}).
The tranformations (\ref{v(u)1}) and (\ref{v(u)2}) may then be written in the
compact form 
\be
\exp\left(-\kappa \vt^-\right) = 
\frac{1}{2}\,\exp\left(-\kappa \ut^-\right)\,(u^-_H-\ut^-).
\label{vu}
\ee
From this it follows that the complex points $\vt^-$ and $\vt^-+i2\pi/\kappa$ 
represent the same event in spacetime.

\begin{figure}
\centerline{\epsfysize=2.5truein\epsfbox{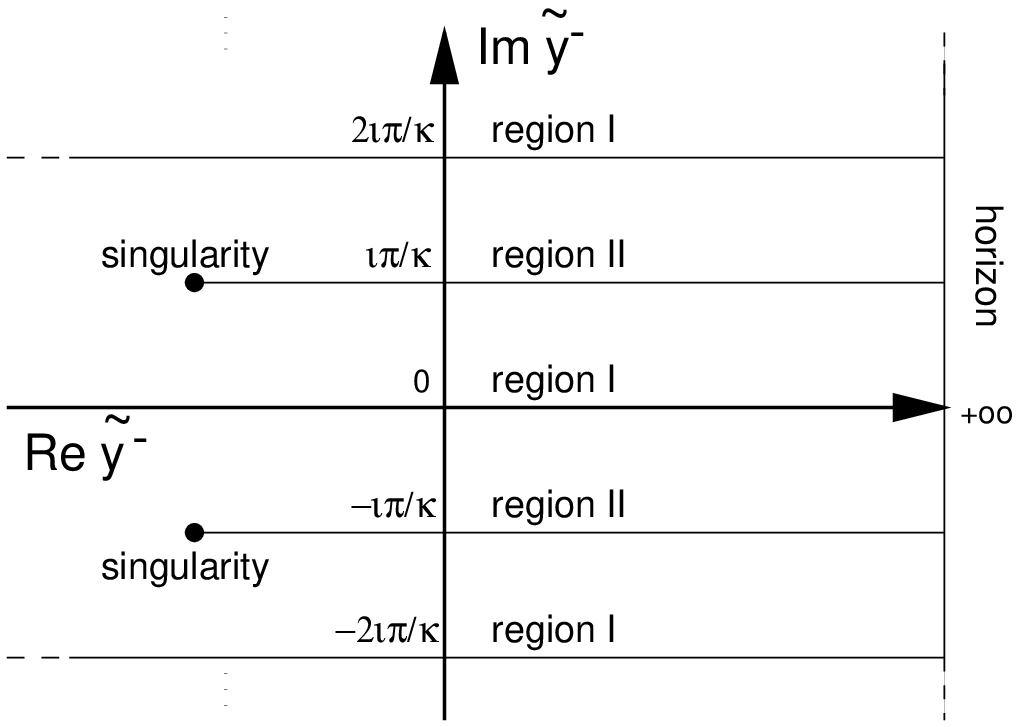}}
\vspace{5mm}
\caption{ A section of the tortoise complex covering space $\Vct$ along the 
$\vt^-$ coordinate.}
\label{fig:schconf}
\end{figure}

\begin{figure}
\centerline{\epsfysize=1.6truein\epsfbox{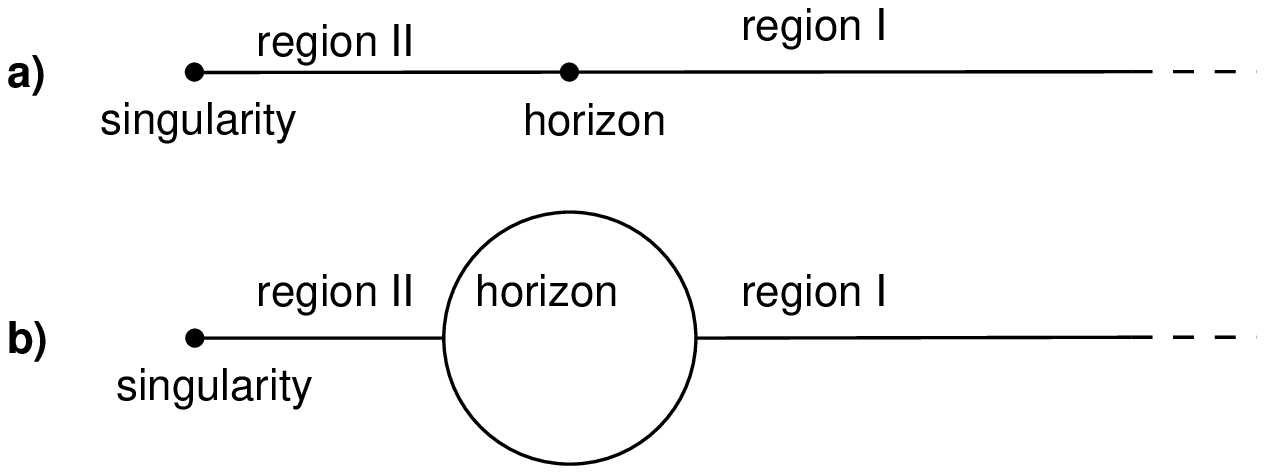}}
\vspace{5mm}
\caption{A section of the configuration spaces a) $\Uc$ and b) $\Vc$
along the $u^-$ and $v^-$ coordinates respectively. }
\label{fig:schtopo}
\end{figure}

Since the configuration space in tortoise coordinates follows directly
form the transformation (\ref{vu}), or from the transformations
(\ref{u(v)R1}) and (\ref{u(v)R2}) close to the event-horizon, one can
see that the homotopy of the configuration space $\Vc$ in the $v$
coordinates is analogous to that found above for the $y$
coordinates. The complex configuration space $\Vct$ is given by
\begin{eqnarray}
\Vct = \{\,(\vt^+,\vt^-,\Omt)\in\R\times
\left[\bigcup_{\mu\,\in\,\Z} A_{\mu/2}\cup B\right]\times\s^2 
\mbox{ such that } r(v^+,v^-)>0\,\},
\end{eqnarray}
where the sets $A_\mu$ and $B$ are defined after Eq.~(\ref{AB}), see
Fig.~\ref{fig:schconf}. The topology of the horizon in tortoise complex
configuration space is circular, it is given by
$\R\times\s^1_{1/\kappa}\times\s^2$ in the $(v^+,v^-,\Omega)$
coordinates, see Fig.~\ref{fig:schtopo}.

Since the configuration space $\Vct$ is invariant under the action of the 
transformations  $\vt^-\mapsto\vt^-+i2\pi\nu/\kappa$ where $\nu\in\Z$,
the propagator $\Kt_v$ satisfies
\be
\Kt_v\left(\vt_i;\vt^+_f,\vt_f^-+i2\pi\nu/\kappa,\Omt_f;s\right) = 
\Kt_v\left(\vt_i^+,\vt_i^--i2\pi\nu/\kappa,\Omt_i;\vt_f;s\right).
\ee
From Eq.~(\ref{sumnu}) we obtain furthermore
\be
K^{in}_v\left(v_i;v_f;s\,\right)=\sum_{\nu\in\Z} 
K^{out}_v \left(v_i;v^+_f, v^-_f+i2\pi\nu/\kappa,\Omega_f;s\,\right).
\label{Sfinalbis}
\ee
These expressions are true everywhere.
However, contrary to the Gui spacetime, the propagator $K^{out}_v$ is
{\it not} a free propagator in the present case.
The propagator $K^{in}_v$ is obviously periodic in the imaginary
direction $v^-$,
\be
K^{in}_v\left(v_i;v^+_f, v^-_f,\Omega_f;s\,\right) =
K^{in}_v\left(v_i;v^+_f, v^-_f+i2\pi\nu/\kappa,\Omega_f;s\,\right).
\ee
We interpret these results as indicating that an exact calculation of the 
two-point function in the Unruh vacuum for the collapsing black hole
should include a sum over paths that probe the non-trivial topology of
the tortoise coordinates.

\section{Conclusions}
 
We have shown that if tortoise type coordinates are extended across
an event horizon, this extension can be carried out in a denumerable
infinite number of ways, and that this construction may be used to
consider a Lorentzian path integral in tortoise coordinates which
includes a sum over paths that cross the horizon. Computing such a
path integral in the simple case of a non-static version of the Gui
spacetime, we find that the propagator given by this choice of paths
to sum over is a propagator which is periodic in the null coordinate
transverse to the horizon, and whose energy-momentum tensor exhibits a
thermal flux. A similar propagator has been obtained in a collapsing
black-hole spacetime, and we conjecture that the corresponding
propagator defines the Unruh vacuum.
 
It is interesting to note that while in Kruskal coordinates, the main
contribution to the path integral comes from the geodesics joining the
two endpoints, in tortoise coordinates, the contributions of paths
crossing the horizon must also be considered.  In the non-static Gui
spacetime, the quantum particle may cross the horizon an arbitrary
number of times, even though the regions $I$ and $I\!I$ are
classically causally disconnected.

We conclude that the homotopic properties of the configuration space
are not an intrinsic feature of a spacetime, but depend on the set of
coordinates chosen to cover it, and may be hidden in the complex
nature of coordinate relations across horizons.  These properties have
no physical consequences for the classical motion of a particle.
However, in a quantum mechanical framework, a particle may tunnel
across a horizon, and the topology of the whole configuration space is
then physically relevant.  We conjecture that the thermal nature of
event horizons can be derived by using the fact that there is a
denumerably infinite number of ways for a particle to tunnel through
the horizon.  These are defined by the winding number of the particle
paths around the horizon in the tortoise configuration space.

\acknowledgments

We would like to thank J.~J.~Halliwell and M.~B.~Mensky for helpful 
conversations.


\begin{thebibliography}{}
  
\bibitem{FHS} R.~P.~Feynman and A.~R.~Hibbs, {\it Quantum mechanics
    and path integrals} (McGraw-Hill Book Co., New York, 1965);
  L.~S.~Schulman, {\it Techniques and application of path integrals}
  (John Wiley and Sons, New York, 1981).
  
\bibitem{WMN} C.~DeWitt-Morette, A.~Maheshwari and B.~Nelson,
  Phys.~Rep.~50, 255 (1979).
  
\bibitem{SD} L.~S.~Schulman, Phys.~Rev.~176, 1558 (1968);
  J.~Math.~Phys.~12, 304 (1971); J.~S.~Dowker,
  J.~Phys.~A:~Gen.~Phys.~5, 936 (1972).

\bibitem{kms} R. Kubo, J.~Phys.~Soc.~Jpn.~12, 570 (1957); 
  P.~C.~Martin and J. Schwinger, Phys.~Rev.~115, 1342 (1959).
  
\bibitem{troost} W.~Troost and H.~Van Dam, Phys.~Lett.~B 71, 149 (1977); 
  Nucl.~Phys.~B 152, 442 (1979).

\bibitem{gui} Y.-X.~Gui, Phys.~Rev.~D 42, 1988 (1990); 
  Phys.~Rev.~D 46, 1869 (1992).

\bibitem{HH} J.~B.~Hartle and S.~W.~Hawking, Phys.~Rev.~D 13, 2188 (1976).
  
\bibitem{Ha} S.~W.~Hawking, Commun.~Math.~Phys.~43, 199 (1975).
  
\bibitem{LL} M.~Lachi\`eze-Rey and J.-P.~Luminet, Phys.~Rep.~254, 136 (1995).
  
\bibitem{BP} J.~D.~Bekenstein and L.~Parker, Phys.~Rev.~D 23, 2850 (1981); 
  L.~Parker, in {\it Recent developments in gravitation,
  Carg\`ese 1978}, edited by M.~L\'evy and S.~Deser (Plenum Press,
  New York, 1979).

\bibitem{Ma} J.~S.~Dowker, J.~Phys.~A 5, 936 (1972);
M.~S.~Marinov, Phys.~Rep.~60, 1 (1980).

\bibitem{FV} F.~Vendrell, Helv.~Phys.~Acta 70, 598 (1997).

\bibitem{HO} J.~J.~Halliwell and M.~E.~Ortiz, Phys.~Rev.~D 48, 748 (1993).
  
\bibitem{Sy} J.~L.~Synge, Proc.~Roy.~Irish Acad.~59 A, 1 (1957).

\end{thebibliography}
\end{document}